\documentstyle[preprint,epsfig,eqsecnum,aps]{revtex}

\title{Effects of the secondary decays on the isotopic thermometers}
\author{Al. H. Raduta and Ad. R. Raduta}
\address{National Institute of Physics and Nuclear Engineering,\\ 
  Bucharest, POB MG-6, Romania}
\begin{document}
\draft
\maketitle
\begin{abstract}
  The sharp microcanonical multifragmentation model from [Al. H. Raduta and
  Ad. R. Raduta, Phys. Rev. C {\bf 55}, 1344 (1997); Phys. Rev. C, in press]
  is employed for evaluating the nuclear caloric curve predictions of nine
  isotopic thermometers for three representative nuclei. Evaluations are
  performed for both primary decay and asymptotic stages. Effects of the
  secondary decays on the primary decay caloric curves are evidenced and
  discussed. In both cases a dispersive character of the isotopic caloric
  curves with increasing the source excitation energy is observed. A procedure
  of calibrating the isotopic thermometers on the microcanonical predictions
  for both primary decay and asymptotic stages is proposed. The resulting set
  of calibrating parameters for each thermometer is independent on the source
  size, on its excitation energy and, in the case of the primary decay, on the
  freeze-out radius assumption.
\end{abstract}
 \newpage Nuclear caloric curve is
presently one of the most studied subjects in nuclear multifragmentation
research. The reason lay in the resemblance of the nucleon-nucleon interaction
with the van der Waals forces which motivates the expectation of a liquid-gas
phase transition in nuclear matter.  While statistical multifragmentation
models (SMM \cite{Bondorf} and MMMC \cite{Gross}) predicted caloric curves
with transition-like regions since 1985, the first experimental evaluation of
the nuclear caloric curve was performed ten years later by the ALADIN
collaboration \cite{Pocho}. A wide plateau interrupting the monotonical
increase of the caloric curve at 5 MeV temperature was evidenced with that
occasion. This was interpreted as a sign of liquid-gas phase transition.
Experiments of INDRA \cite{Indra} and EOS \cite{Eos} performed on other
reactions followed showing sightly different results.  Two years latter,
papers of the ALADIN collaboration \cite{Xi,Trautmann} announced a
reevaluation of the neutron kinetic energies and of the energies of the
spectator parts resulting in a modification of the ALADIN caloric curve from
1995. Reasons for such discrepancies lay in both the different employed
reactions leading to different sequences of equilibrated sources ($E^*(A)$)
and, as shown above, in the degree of precision of the experimental
measurements. The question is to what extent the experimentally measured
caloric curves relate to the {\em real} one. Apart from the nonequilibrium
parts of the excitation energies of the nuclear sources which have to be
properly removed, one has to use a temperature formula which accounts for the
finite size effects which manifest in nuclear systems.

In this respect, is widely used the Albergo isotopic formula \cite{Albergo},
expressed as a double ratio of isotopic yields:
\begin{equation}
  T_{12/34}=\Delta B/ \ln \{s \left[ Y(A_1,Z_1)/Y(A_2,Z_2)\right]/
  \left[ Y(A_3,Z_3)/Y(A_4,Z_4) \right] \},
\label{eq:tizo}
\end{equation}
where $\Delta B=\left[ B(A_1,Z_1)-B(A_2,Z_2)\right]-
\left[B(A_3,Z_3)-B(A_4,Z_4) \right]$, $B(A_i,Z_i)$ and $Y(A_i,Z_i)$ are
respectively the binding energy and the yield corresponding to the isotope
$(A_i,Z_i)$, $s$ is a statistical factor deduced from the ground state spin
and masses of the isotopes and, finally, the isotope pairs (1,2) and (3,4)
must differ by the same number of neutrons and/or protons. Being deduced on
grandcanonical grounds, formula (\ref{eq:tizo}) does not account for finite
size effects.  Pochodzalla {\it et al.} \cite{Pocho} tried to remediate this
drawback by including a calibrating factor $f_T=1.2$ chosen as to average the
results of the QSM, GEMINI and MMMC models. Obviously, the efficiency of this
method depends on the compatibility of the employed model with the physical
phenomenon. Due to the finite dimensions of the systems currently involved,
the microcanonical ensemble is the best solution.  Moreover, one has to take
into account the process of secondary decays from the primary decay excited
fragments which modify the yields entering formula (\ref{eq:tizo}).

The calibration of seven isotopic thermometers on the prediction of the sharp
microcanonical model from Ref. \cite{Noi1} was described in \cite{Temp3}. 
There, the
Albergo thermometers were {\em only} microcanonically corrected. In order to
allow a direct application of the thermometers on the experimental yields one
has to apply a {\em second} correction for eliminating the effects of the
secondary decays. This is now possible with the new version \cite{Temp4} of the
sharp microcanonical multifragmentation model \cite{Noi1}. In comparison with 
its initial formulation, in Ref. \cite{Temp4} the microcanonical model is 
refined and
improved by considering the experimental energy levels for fragments with $A
\le 6$ (instead of the level density Fermi gas formula) and by including the
secondary decay stage.  Using this version, a very good agreement was obtained
in Ref. \cite{Temp4} between the predicted uncorrected HeLi caloric curve
and the experimental (uncorrected) HeLi caloric curve of ALADIN from 1997
\cite{Xi,Trautmann}.

The present paper intends to use the new version \cite{Temp4} of the sharp
microcanonical model \cite{Noi1} in order to analyze the effects of the
secondary decays on the primary decay isotopic caloric curve and to perform a
microcanonical calibration of nine isotopic thermometers for both primary
decay and asymptotic stages. Also, a study is made concerning the sensitivity
of the resulting parametrization to the freeze-out radius assumption.

\begin{figure}
  \begin{center}
  \epsfig{file=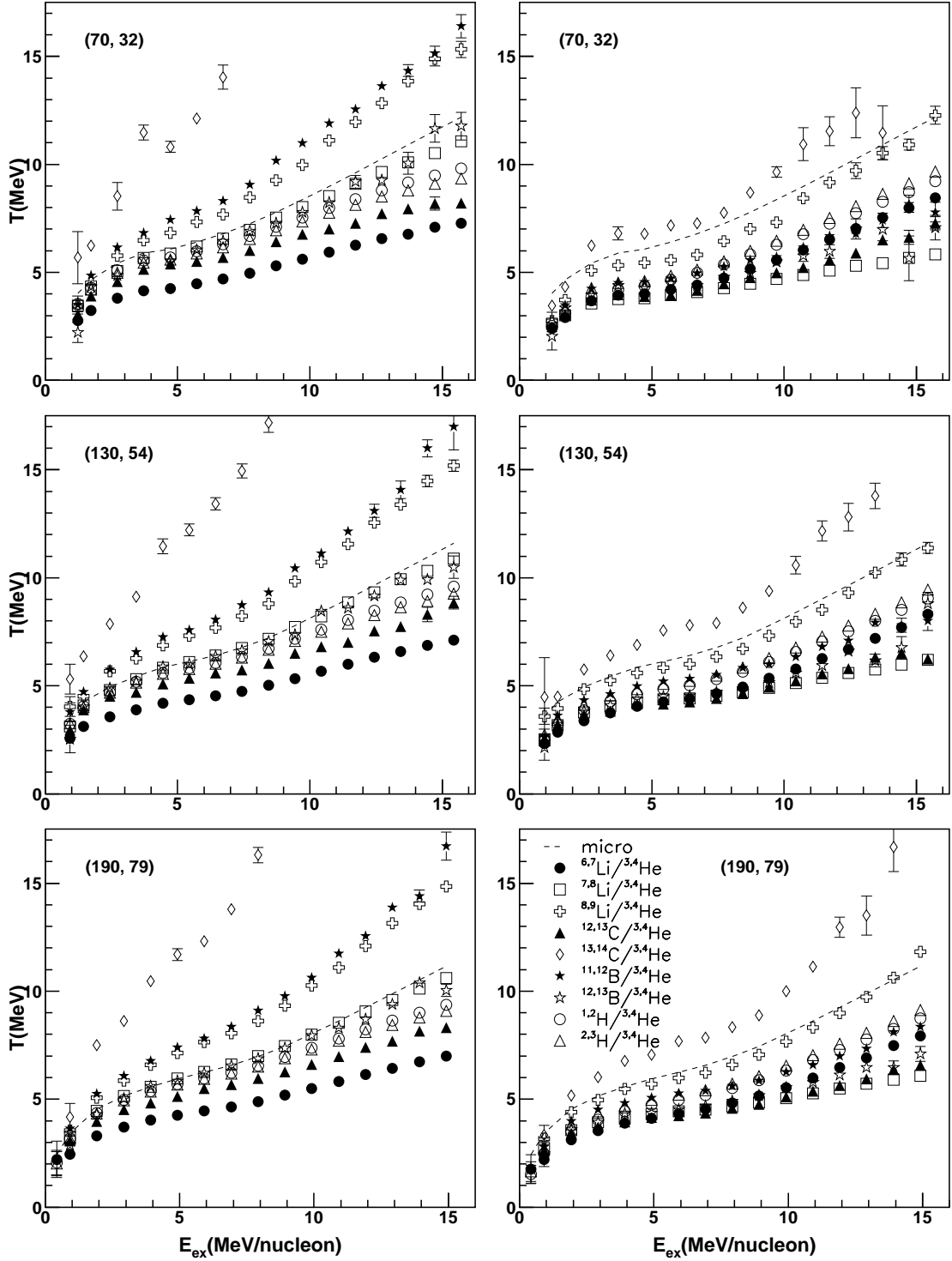,height=19.5cm,angle=0}
  \end{center}
  \caption{Caloric curves corresponding to nine isotopic thermometers
    evaluated for three nuclear sources with the freeze-out radius R=2.5
    A$^{1/3}$ fm in both primary decay (left column) and asymptotic (right
    column) stages. The microcanonical caloric curve is represented by a
    dashed line.}
\label{fig:1}
\end{figure}

Isotopic caloric curve evaluations are presented in Fig. \ref{fig:1} for nine
isotopic thermometers: $^{6,7}$Li/$^{3,4}$He, $^{7,8}$Li/$^{3,4}$He,
$^{8,9}$Li/$^{3,4}$He, $^{12,13}$C/$^{3,4}$He,
$^{13,14}$C/$^{3,4}$He, $^{11,12}$B/$^{3,4}$He, $^{12,13}$B/$^{3,4}$He,
$^{1,2}$H/$^{3,4}$He, $^{2,3}$H/$^{3,4}$He together with the microcanonical
caloric curve, in the case of three representative source nuclei (70,32),
(130,54) and (190,79) and a freeze-out radius parameter of 2.5 $A^{1/3}$ fm.
Calculations have been performed for both primary decay stage (left column)
and asymptotic stage (right column).  In all cases the isotopic curves show a
dispersive character growing with the increase of the source excitation
energy. An interesting effect is the strong diminish of this dispersion after
the secondary decays (see Fig. \ref{fig:1}, right column).  When analyzing the
asymptotic curve this effect may induce the {\em exaggerate} impression that
the uncorrected Albergo isotopic thermometers give quite close results and
consequently, the finite size effects affecting formula (\ref{eq:tizo}) are
small. In fact, secondary decays can bring some isotopic caloric curves quite
close to the microcanonical ones in spite of the great deviation between the
above mentioned curves appearing in the primary decay stage (e.g. the case 
of the $^{13,14}$C/$^{3,4}$He curve).

\begin{figure}
  \begin{center}
  \epsfig{file=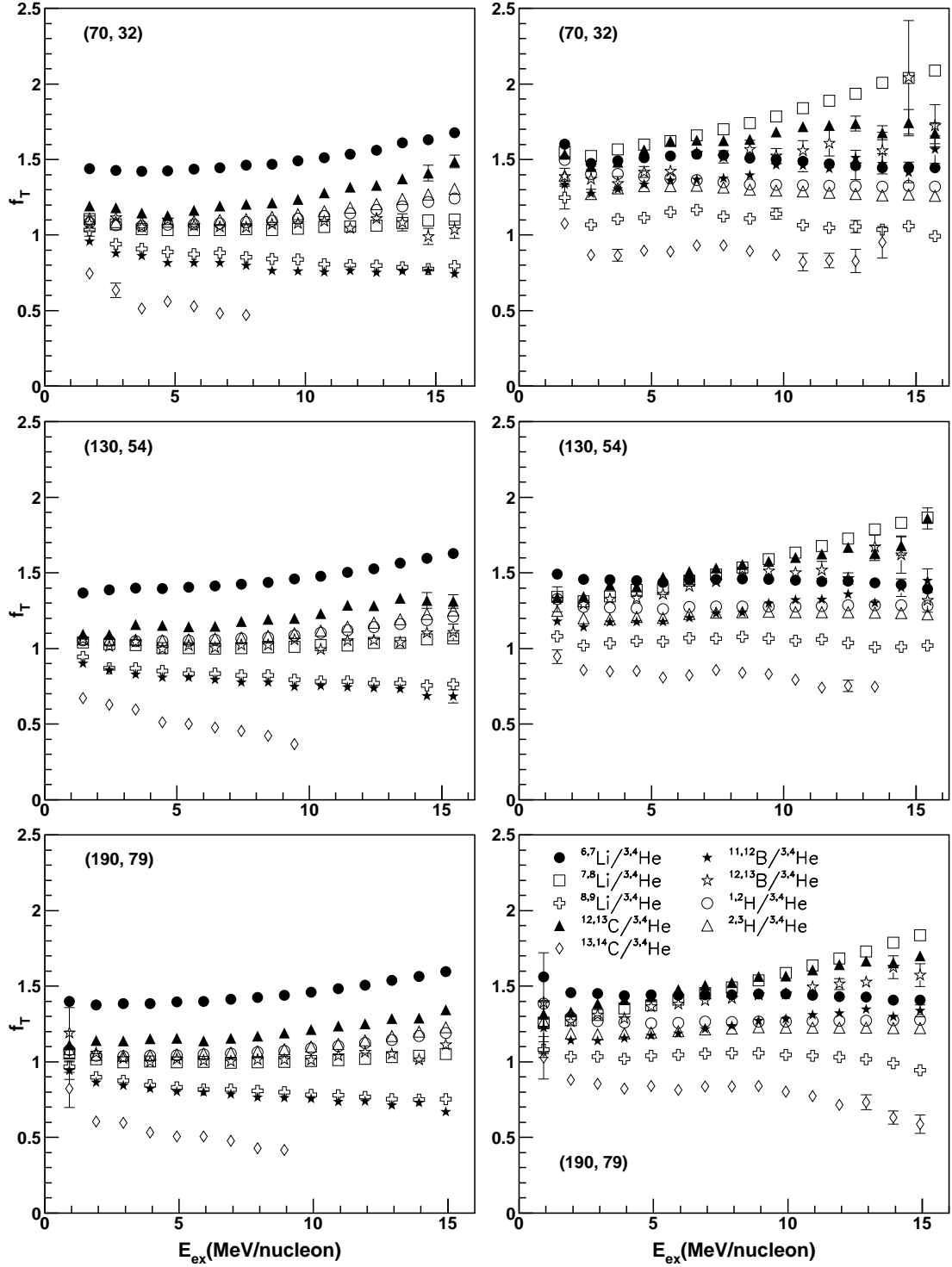,height=19.5cm,angle=0}
  \end{center}
  \caption{Calibration factors corresponding to nine isotopic thermometers
    evaluated for three nuclear sources with the freeze-out radius R=2.5
    A$^{1/3}$ fm in both primary decay (left column) and asymptotic (right
    column) stages.}
\label{fig:2}
\end{figure}

In order to calibrate formula (\ref{eq:tizo}) on the microcanonical
predictions for both primary decay and asymptotic stages we apply the method
described in Ref. \cite{Temp3}. First, the calibrating factors
$f_T=T_{micro}/T_{iso}$ (where $T_{micro}$ denotes the microcanonical
temperature and $T_{iso}$ the corresponding isotopic temperature) 
corresponding to each temperature point from Fig. \ref{fig:1} are 
represented in Fig. \ref{fig:2}. In all cases, the $f_T(E_{ex})$ curves are
almost straight so one can use linear expressions for fitting them:
\begin{equation}
  f_T(E_{ex})=a~E_{ex}+b,
  \label{eq:ft}
\end{equation}
where $a$ and $b$ are calibrating parameters. A set of calibrating parameters
$a$, $b$ was derived for each considered thermometer for both the primary
decay and the asymptotic situations. The values of these parameters are given
in Table \ref{table:1} for the case of the primary decay and in Table
\ref{table:2} for the asymptotic case.

\begin{figure}
  \begin{center}
  \epsfig{file=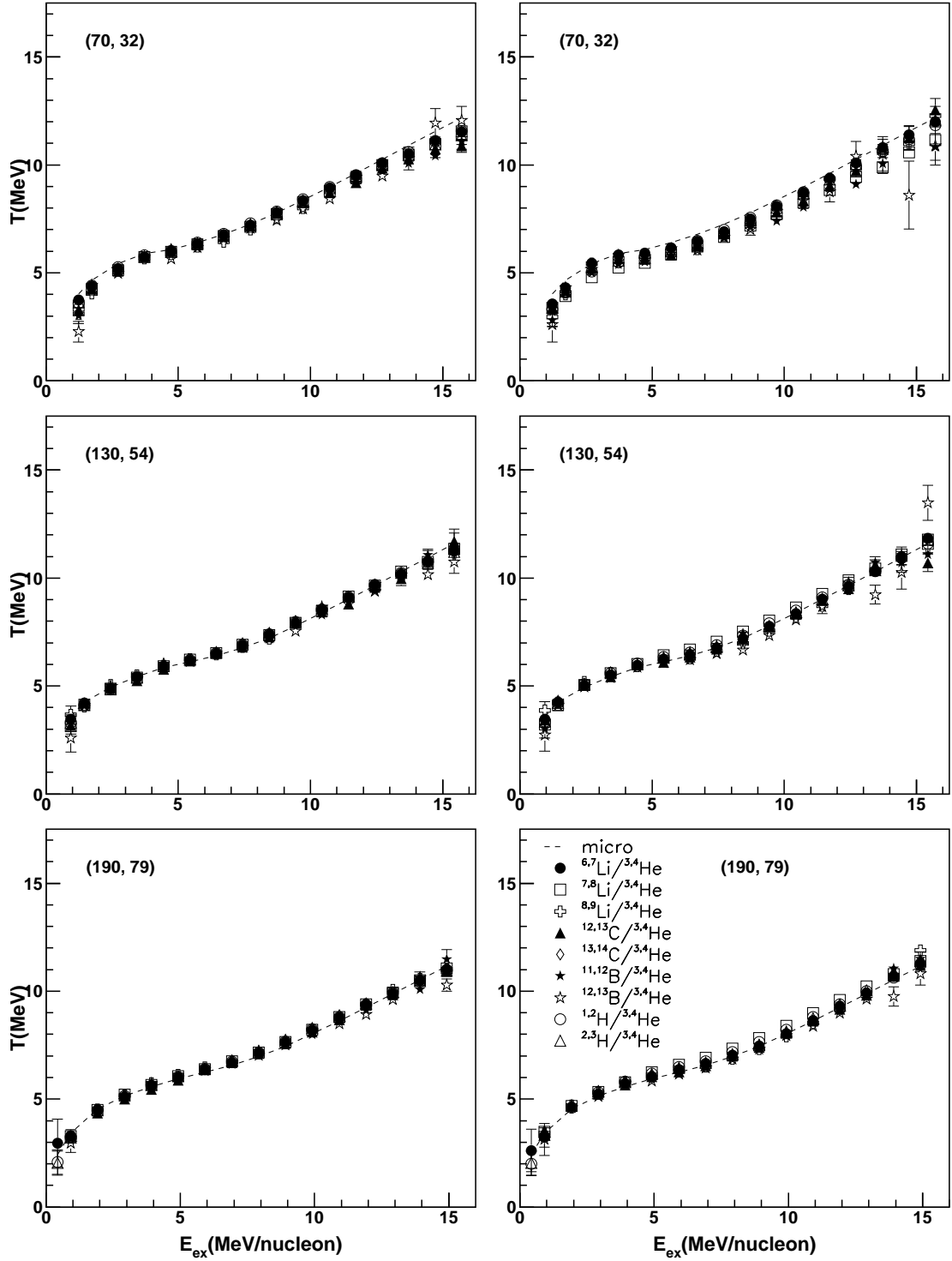,height=19.5cm,angle=0}
  \end{center}
  \caption{The isotopic caloric curves from Fig. \ref{fig:1} calibrated by
    means of equation (\ref{eq:tcorr}) and the parameters from Tables
    \ref{table:1} and \ref{table:2}. The microcanonical caloric curve is
    represented by a dashed line.}
\label{fig:3}
\end{figure}

In order to check the efficiency of the obtained parametrization, the isotopic
caloric curves corresponding to both primary decay and asymptotic situations
(Fig. \ref{fig:1}) are adjusted using the obtained calibration parameters
according to the relation:
\begin{equation}
  T_{iso}^{corr}(E_{ex})=T_{iso}(E_{ex})~f_T(E_{ex}).
  \label{eq:tcorr}
\end{equation}
The result is the overlapping between all the isotopic caloric curves and the
microcanonical one for both the primary decay stage and the asymptotic stage
presented in Fig. \ref{fig:3}. In conclusion, $f_T(E_{ex})$ is perfectly
applicable for calibrating formula (\ref{eq:tizo}) correctly. It is remarkable
that the obtained parameters do not depend on the source size or its
excitation energy. This makes them quite applicable on experimental data where
the source size is strongly dependent on its excitation energy. While the
parametrization corresponding to the primary decay case (Table
\ref{table:1}) needs a further correction as to erase the effects of the
secondary decays, the parameterization corresponding to the asymptotic
situation (Table \ref{table:2}) is directly applicable on the experimental
isotopic yields.

\begin{figure}
  \begin{center}
  \epsfig{file=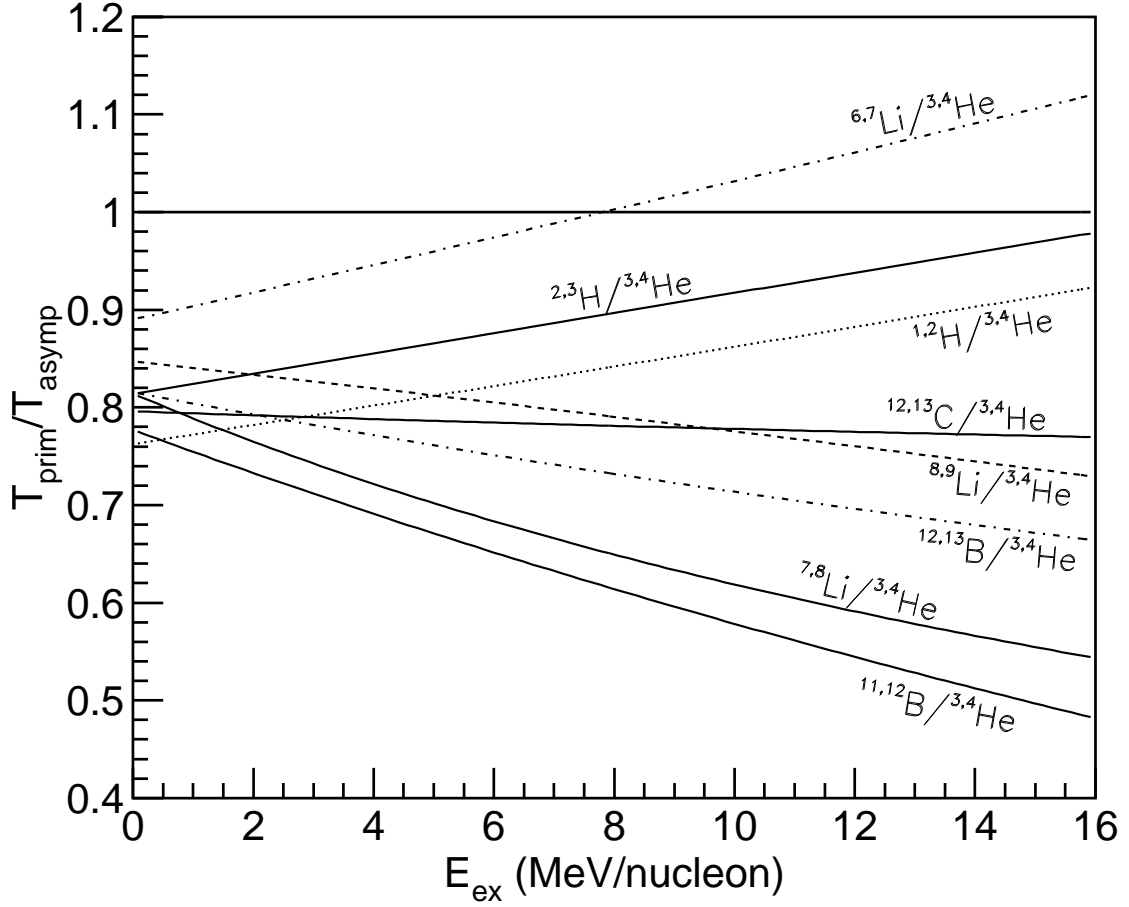,height=12cm,angle=0}
  \end{center}
  \caption{Ratios between primary decay isotopic temperature and 
    asymptotic isotopic temperature versus excitation energy approximated by
    the ratio $f_T^{asymp}(E_{ex})/f_T^{prim}(E_{ex})$ with $f_T(E_{ex})$
    given by equation (\ref{eq:ft}) and the parameters $a$ and $b$ taken from
    Table \ref{table:1}.}
\label{fig:4}
\end{figure}

The modifications brought by the secondary decays to the primary decay
isotopic caloric curve are clearly evidenced in Fig. \ref{fig:4}. There it is
represented the ratio between the primary decay caloric curves and the
asymptotic caloric curve for all cases except $^{13,14}$C/$^{3,4}$He. For the
clarity of the figure, this ratio can be very well approached by the ratio of
the factors $f_T(E_{ex})$ corresponding to the asymptotic situation and those
corresponding the the primary decay situation,
$f_T^{asymp}(E_{ex})/f_T^{prim}(E_{ex})$. Except for one situation
($^{6,7}$Li/$^{3,4}$He), the curves after the secondary decays are lower
than those corresponding to the primary decay on the entire considered domain
of excitation energy. Also, the deviation of the asymptotic curves from the
primary decay curves increases in five cases ($^{7,8}$Li/$^{3,4}$He,
$^{8,9}$Li/$^{3,4}$He, $^{12,13}$C/$^{3,4}$He, $^{11,12}$B/$^{3,4}$He,
$^{12,13}$B/$^{3,4}$He) and decreases in other three cases
($^{6,7}$C/$^{3,4}$He, $^{1,2}$H/$^{3,4}$He, $^{2,3}$H/$^{3,4}$He).

\begin{figure}
  \begin{center}
  \epsfig{file=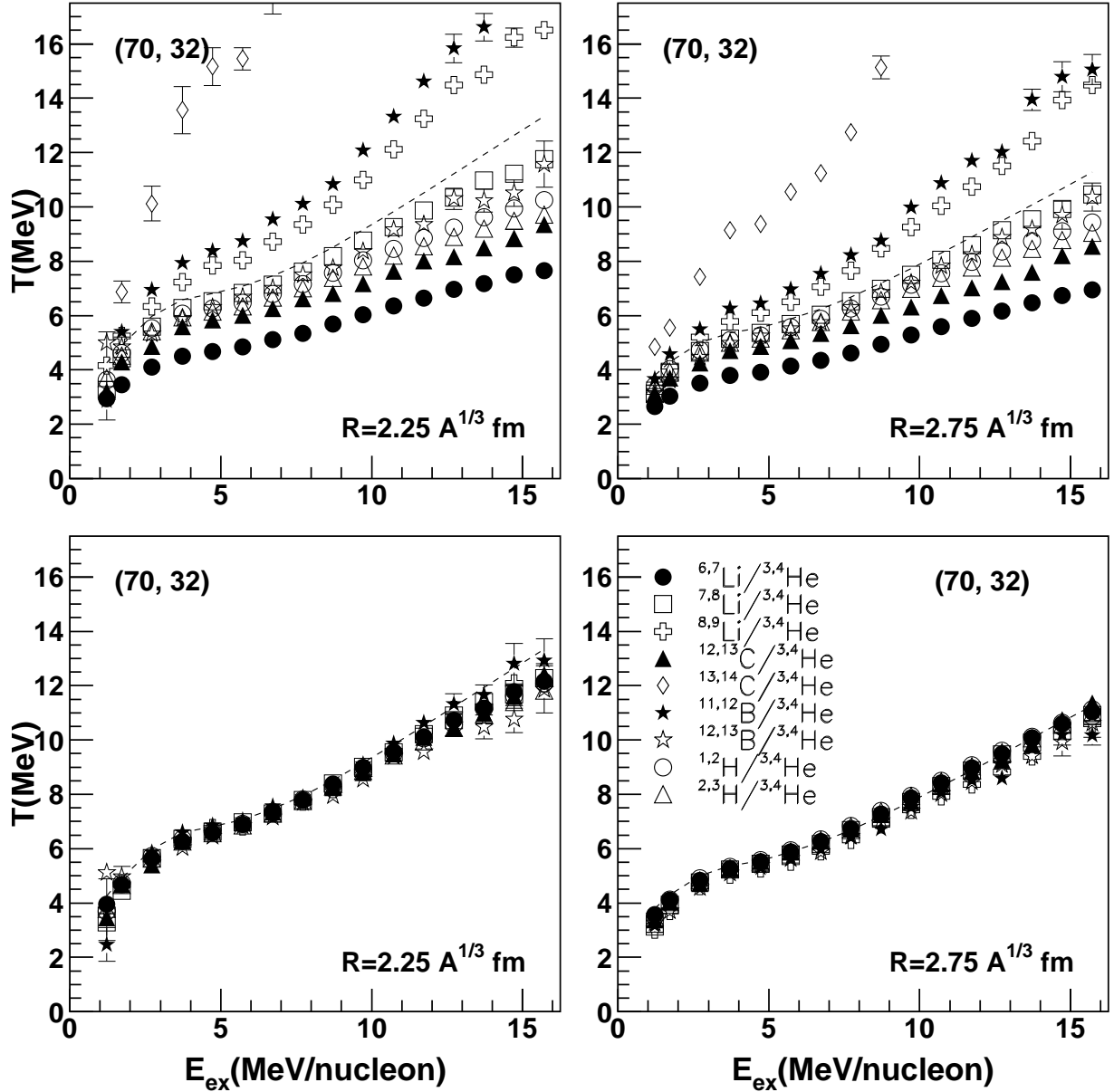,height=16cm,angle=0}
  \end{center}
  \caption{Caloric curves corresponding to nine isotopic thermometers 
    evaluated for the nuclear source (70,32) and the freeze-out radii 2.25
    A$^{1/3}$ fm and 2.75 A$^{1/3}$ fm (upper side) and the same caloric
    curves calibrated by means of the parameters taken from Tables
    \ref{table:1} and \ref{table:2} (lower side). The microcanonical caloric
    curve is represented by a dashed line.}
\label{fig:5}
\end{figure}

Finally, a study is made concerning the influence of the freeze-out radius
assumption on the resulting parametrization. To this aim, the isotopic caloric
curves corresponding to the primary decay for the (70,32) source for the
freeze-out radii 2.25 $A^{1/3}$ fm and 2.75 $A^{1/3}$ fm are plotted in the
left column of Fig. \ref{fig:5}. The overall aspect is maintained except the
position of the plateau-like region. In the left column of Fig. \ref{fig:5}
the caloric curves corresponding to each considered freeze-out radius,
adjusted according to equation (\ref{eq:tcorr}) using the parameters listed in
Table \ref{table:1} are represented. Remarkably, the corrected curves overlap
over the microcanonical ones. This means that the parametrization
corresponding to the primary decay (Table \ref{table:1}) is also independent
to the freeze-out radius assumption. A similar study concerning the asymptotic
isotopic caloric curves using the parameters from Table \ref{table:2} showed
the same tendency of overlapping between the isotopic curves but a small
global deviation from the microcanonical curve was present for the $2.25
A^{1/3}$ case.

Summarizing, isotopic caloric curves have been evaluated for nine isotopic
thermometers in both primary decay and asymptotic stages. In both cases a
dispersive character of the isotopic caloric curve monotonically increasing
with the increase of the excitation energy is evidenced. In the asymptotic
stage this dispersion is strongly diminished. A set of microcanonical
calibrating parameters was deduced for each of the considered thermometers for
both primary decay and asymptotic situations.  Remarkably, these parameters
are independent on the dimension of the source, on its excitation energy and,
in the primary decay stage, on the freeze-out radius assumption. Finally, a
study was made concerning the modifications brought by the secondary decays to
the primary decay isotopic caloric curves.

\begin{table}
\begin{tabular}{ccc}
Thermometer          &a(nucleon $\times$ MeV$^{-1}$)&b\\
\hline
$^{6,7}$Li/$^{3,4}$He& 0.016510             & 1.328440\\
$^{7,8}$Li/$^{3,4}$He& 0.002526             & 1.004070\\
$^{8,9}$Li/$^{3,4}$He& -0.011714             & 0.916471\\
$^{12,13}$C/$^{3,4}$He&0.016886             & 1.061580\\
$^{11,12}$B/$^{3,4}$He&-0.013010            & 0.880336\\
$^{12,13}$B/$^{3,4}$He&0.000098             & 1.023200\\
$^{1,2}$H/$^{3,4}$He  &0.011133             & 1.004980\\
$^{2,3}$H/$^{3,4}$He  &0.013426             & 1.004870\\
\end{tabular}
\caption{Calibrating parameters corresponding to the primary decay stage.}
\label{table:1}
\end{table}

\begin{table}
\begin{tabular}{ccc}
Thermometer          &a(nucleon $\times$ MeV$^{-1}$)&b\\
\hline
$^{6,7}$Li/$^{3,4}$He& -0.004475            & 1.492390\\
$^{7,8}$Li/$^{3,4}$He&  0.043046            & 1.233270\\
$^{8,9}$Li/$^{3,4}$He& -0.005079            & 1.081550\\
$^{12,13}$C/$^{3,4}$He& 0.024806            & 1.333520\\
$^{11,12}$B/$^{3,4}$He& 0.016407            & 1.133040\\
$^{12,13}$B/$^{3,4}$He& 0.018067            & 1.254630\\
$^{1,2}$H/$^{3,4}$He  & -0.002311           & 1.318080\\
$^{2,3}$H/$^{3,4}$He  &  0.000675           & 1.235100\\
\end{tabular}
\caption{Calibrating parameters corresponding to the asymptotic stage.}
\label{table:2}
\end{table}

\end{document}